\documentclass[aps,twocolumn,showpacs]{revtex4}
\usepackage{amsmath}%
\usepackage{amsthm,amsmath,amssymb}
\usepackage{comment}
\usepackage{mathrsfs}
\usepackage{graphicx}
\usepackage{dcolumn}
\usepackage{bm}
\usepackage{color, soul}
\usepackage{color, xcolor}
\usepackage[utf8]{inputenc}
\usepackage[T1]{fontenc}
\usepackage{mathptmx}
\usepackage{multirow}
\usepackage{makecell}
\usepackage{setspace}
\usepackage{footnote}
\usepackage{hyperref}
\hypersetup{hypertex=true,
	colorlinks=true,
	linkcolor=blue,
	anchorcolor=blue,
	citecolor=blue}

\begin{document}
\title{Nonlinear Faraday magneto-optic effects in a helically wound optical fiber}

\author{Peng Gao$^{1,2}$}
\author{Bin Sun$^{2}$}
\author{Jie Liu$^{2}$}\email{jliu@gscaep.ac.cn}

\address{$^1$School of Physics, Xidian University, Xi’an 710071, China}
\address{$^2$Graduate School, China Academy of Engineering Physics, Beijing 100193, China}

\begin{abstract}
We thoroughly investigate the Faraday magneto-optical effects in a helically wound nonlinear optical fiber.
We find the emergence of an additional rotation angle proportional to the optical intensity, arised from the nonlinear corrections to both the Verdet constant and the fiber torsion.
By analyzing an oscillator model describing the electron motions in the fiber medium, we can obtain the third-order susceptibility in the presence of the magnetic field.
According to the Maxwell's equations and the minimal coupling principle in the metric expression of the curved space, we derive the propagation equations of light in a helically wound optical fiber.
Finally, we have obtained the analytic expressions of Faraday rotation angle for both linearly and elliptically polarized lights, which explicitly indicates an important nonlinear correction on the Faraday rotation angles.
Possible experimental observations and some implications of our theoretical findings are discussed.
\end{abstract}

\maketitle

\section{Introduction}

The Faraday magneto-optical effect, or the Faraday rotation, is a fundamental and significant physical phenomenon. It describes the rotation of a light beam's polarization when passing through a medium under the influence of a magnetic field \cite{Faraday-1846}.
This circular-birefringence phenomenon induced by external fields has found numerous applications in areas such as optical isolators \cite{Aplet-1964,Takeda-2008,Efimkin-2013}, fiber optic sensing \cite{Mihailovic-2021}, and magnetic field measurement \cite{Budker-2000}.
Under non-resonant conditions, where the optical frequency are considerably smaller than the lowest electronic resonance frequency of the material, its physical mechanism can be effectively explained by the classical oscillator model \cite{Boyd-book,Sato-2022}: electrons rotate in a fixed direction under the Lorentz force, which impacts the medium's response to left- and right-handed circularly polarized lights.
Under near-resonant conditions, quantum theory needs to be introduced to analyze the underlying mechanism \cite{Bennett-1965,Schatz-1969,Sato-2022}, which can explain the distinct nonlinear dependence of Faraday rotation angle on the magnetic field \cite{Kanorsky-1993,Budker-2002,Zhu-2013}.
More interestingly, the rotation of a light beam's polarization can be also generated by the spatial geometric torsion when a light beam propagates through a helically wound optical fiber \cite{Ross-1984,Tomita-1986}.
Underlying mechanism can be understood by the celebrated Berry phase \cite{Berry-1984,Chiao-1986,Berry-1987} associated with parallel transport \cite{Kugler-1988,Cisowski-2022}.

However, conventional Faraday effects are restricted to the situation that the incident light intensity is weak so that the nonlinear response to an intense light is ignored.
With the development of nonlinear optics and strong laser field technologies \cite{Brabec-2000,Toulouse-2005}, some studies recently attempt to consider the nonlinear polarization effects in varied systems, such as multi-level atomic and molecule systems \cite{Yu-1977,Giraud-Cotton-1985}.
The nonlinear Faraday effect in optical fibers might be more interesting, not only due to its possible applications in optical communications \cite{Keiser-book} and sensors \cite{Day-1982,Annovazzi-1992,Lee-2003}, but also because its fundamental physical meanings on the parallel transport and Berry phase in the nonlinear systems \cite{Liu-2010,Liu-book}.

In this paper, we investigate the effects of third-order optical nonlinear response on the rotation angle of light polarization, for a helically wound fiber in the presence of an external magnetic field.
In the non-resonant situation, we obtain the third-order susceptibility of the fiber in the magnetic field, through an electron oscillator model.
From the Maxwell's equations, we derive the evolution equation of optical amplitude with the fiber's arc length, and consequently obtain the rotation angle for both linearly and elliptically polarized lights.
We find that, the Faraday rotation angle exhibits additional terms proportional to the light intensity, which are viewed as the nonlinear corrections to the Verdet constant as well as the helical geometry.
In the absence of magnetic field, the above finding indicates a possible correction on the Berry phase in the nonlinear quantum evolution  \cite{Liu-2010,Liu-book,Zhang-2011}.
Finally, we take a $\rm As_2S_3$ fiber as an example to calculate the nonlinear rotation angle and give its observable condition.

\section{Physical model}

\subsection{Light propagating in a helically wound nonlinear optical fiber}

\begin{figure*}
  \centering
  \includegraphics[width=140mm]{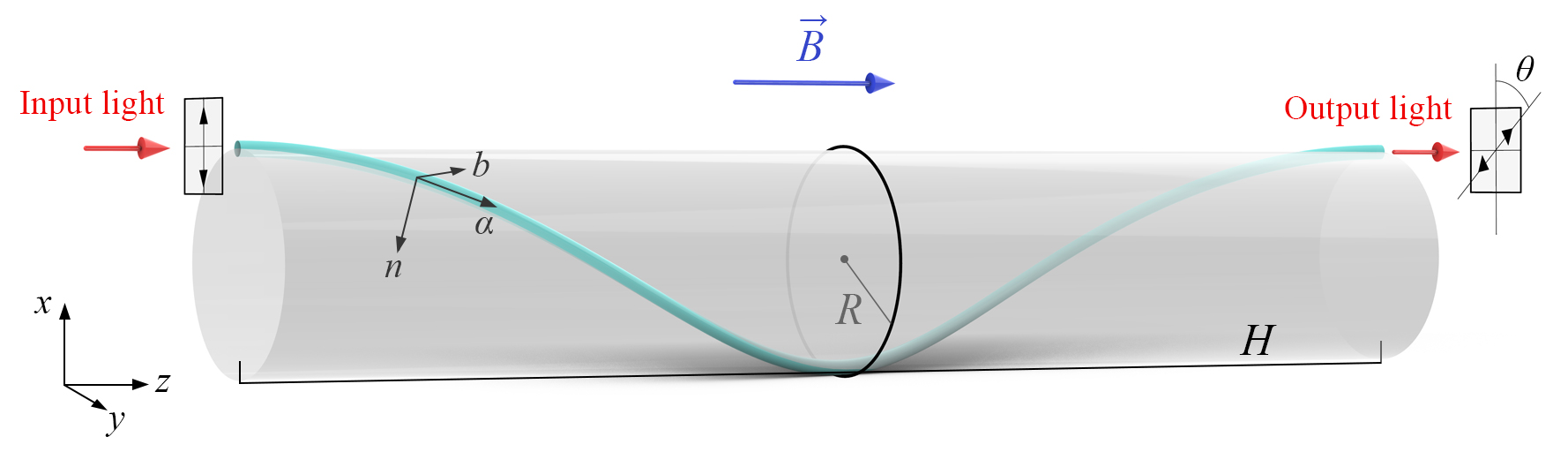}
  \caption{The schematic diagram of the light beam propagating through the helically wound optical fiber in the presence of an external magnetic field $\vec{B}$ is shown below. Here, $R$ and $H$ represent the radius and pitch of the winding, respectively, and $\theta$ is the rotation angle. The basis vectors \(\hat{x},\hat{y},\hat{z}\) constitute a laboratory coordinate system, while \(\hat{n},\hat{b},\hat{\alpha}\) constitute a Frenet coordinate system.}\label{pic-model}
\end{figure*}

We consider a model of a helically wound nonlinear optical fiber, as illustrated in Fig.~\ref{pic-model}.
A segment of optical fiber with length $L$ is wound around a cylinder of radius $R$, which is placed along the $z$ direction.
The pitch of the winding is $H = \sqrt{L^2 - (2\pi R)^2}$.
A magnetic field $\vec{B} = \frac{L}{H}B\,\hat{z}$ is applied in the positive $z$ direction, such that the magnetic field intensity in the tangential direction of the optical fiber is equal to $B$.
A linearly polarized light beam, polarized along the $x$ direction, is input at the left end of the optical fiber.
Due to the presence of the external magnetic field, the helical configuration, and nonlinear effects, the light beam output from the right end will have a polarization direction different from the incident light.
The angle by which the polarization direction differs, denoted by $\theta$, is the rotation angle.
In the absence of magnetic fields and the nonlinearity, this configuration of light propagation has been considered by Refs. \cite{Ross-1984,Tomita-1986,Berry-1987}.

Influenced by the geometric configuration, the light waves in the wound fiber will always propagate tangentially to the fiber. Studying their propagation in the laboratory coordinate system \((\hat{x},\hat{y},\hat{z})\) is inconvenient. Therefore, we will construct a coordinate system dependent on the geometric configuration—the Frenet coordinate system \((\hat{n},\hat{b},\hat{\alpha})\).
The three basis vectors, dependent on the configuration of fiber, have the tangential, normal, and binormal directions of fiber, respectively.
When the origins of the two coordinate systems are the same, dependent on the arc length \(s\), a position vector \(\vec{r}\) can be decomposed in the two coordinate systems as follows:
\begin{eqnarray}
	\begin{split}\label{2a1} \vec{r}=x\,\hat{x}+y\,\hat{y}+z\,\hat{z}=n\,\hat{n}+b\,\hat{b}+\alpha\hat{\alpha},
	\end{split}
\end{eqnarray}
thus, this vector can be expressed in both coordinate systems as $\vec{r}_L=\begin{bmatrix}
			x & y & z
		\end{bmatrix}^{\rm T}$, $\vec{r}_F=\begin{bmatrix}
			n & b & \alpha
		\end{bmatrix}^{\rm T}$,
where subscripts \(L\) and \(F\) represent representations in the laboratory coordinate system and the Frenet coordinate system, respectively.
One can obtain the expressions for the curvature and torsion of the fiber:
\begin{eqnarray}
	\begin{split}\label{kt} &\kappa=\frac{(2\pi)^2R}{L^2},\quad \tau=\frac{2\pi H}{L^2},
	\end{split}
\end{eqnarray}
and the three-dimensional covariant metric tensor:
\begin{eqnarray}
	\begin{split}
		g_{ij}=\begin{bmatrix}
			1 & 0 & G_1 \\
			0 & 1 & G_2 \\
			G_1 & G_2 & G_1^2+G_2^2+G_3^2\\
		\end{bmatrix}_{ij},
	\end{split}
\end{eqnarray}
where $G_1=\kappa\alpha-\tau b, G_2=\tau n, G_3=1-\kappa n$.
The specific calculations about them can be seen in Appendix I.

\subsection{Third-order susceptibility under an external magnetic field}

Before addressing the propagation of polarized lights in the fiber, we need to analyze the nonlinear response of fiber on the light.
Considering that the molecules in fiber belong to the symmetric ones, the nonlinear response is mainly described by the third-order susceptibility \cite{Agrawal-book}.
The classical oscillator model can effectively analyze the polarization properties of the medium under non-resonant conditions, where linear and nonlinear polarizations are distinguished by the harmonic and non-harmonic parts of the electron vibration potential \cite{Boyd-book}.
Here, we will utilize the oscillator model to analyze the influence of the external magnetic field on the third-order susceptibility of the optical fiber, preparing for the subsequent calculation of the Faraday rotation angle.

The equation of motion for the electrons in the optical fiber can be expressed as \cite{Boyd-book}:
\begin{align}\label{move7}
	m_e \ddot{\vec{r}}=-e\vec{E}-(k_0+k_2|\vec{r}|^2)\vec{r}-e\dot{\vec{r}}\times\vec{B},
\end{align}
where $e$ and $m_e$ are the unit charge quantity and mass of electron.
The three terms on the right-hand side represent the electric field force, restoring force, and Lorentz force, respectively.
$k_0$ and $k_2$ are the coefficients of the harmonic and non-harmonic parts of the electron vibration potential.
Assuming the electric field vector always lies in the $x$-$y$ plane, at a fixed position, the electric field components can be expressed as:
\begin{align}\label{Exy61}
	E_j(t)=\frac{1}{2}(A_{j}e^{-i\omega t}+A_{j}^*e^{i\omega t}),
\end{align}
where $j=1,2$ corresponding to the directions of $x$ and $y$, respectively.
Electrons will always move in the $x$-$y$ plane.
Due to the introduction of the non-harmonic potential, the electron's oscillation will contain two parts: the fundamental part with frequency $\omega$ and the third harmonic part with frequency $3\omega$.
Therefore, the displacements in different directions of the electrons can be set as
\begin{align}\label{xy61}
	&r_j(t)=\frac{1}{2}[r_j^{(1)}e^{-i\omega t}+r_j^{(1)*}e^{i\omega t}+r_j^{(3)}e^{-3i\omega t}+r_j^{(3)*}e^{3i\omega t}],
\end{align}
where $r_1=x$ and $r_2=y$.
Considering that the third harmonic part arises from the non-harmonic potential, and the non-harmonic potential is much lower than the harmonic potential, i.e., $k_2|\vec{r}|^2\ll k_0$, we can assume that the amplitude of the third harmonic part is much smaller than that of the fundamental part: $|x^{(3)}|,|y^{(3)}|\ll |x^{(1)}|,|y^{(1)}|$ \cite{Boyd-book}.

Firstly, substituting Eq. (\ref{xy61}) into Eq. (\ref{move7}), retaining terms containing $e^{-i\omega t}$, and ignoring terms containing $k_2$ and $x^{(3)},y^{(3)}$, 
we can solve the amplitudes of the fundamental part $x^{(1)},y^{(1)}$ [see Eq. (\ref{x1y1}) in Appendix II], which are the linear functions about $A_x$ and $A_y$.
We know that the linear polarization vector $\vec{P}^{(L)}$ only contains the fundamental wave with frequency $\omega$, and its $i$-th component has two different expressions:
\begin{align}\label{PL}
  P^{(L)}_i=\epsilon_0\chi^{(1)}_{ij}E_j=-Ner^{(1)}_i,
\end{align}
where $\chi^{(1)}$ is the first-order susceptibility, and $r^{(1)}_1=x^{(1)}$, $r^{(1)}_2=y^{(1)}$.
From Eq. (\ref{PL}) and the derived expressions of $x^{(1)},y^{(1)}$, we can obtain the matrix form of the first-order susceptibility:
\begin{align}\label{chi1}
\chi^{(1)}
&=\frac{\chi_0(\omega)}{1-\Omega_c^2}
\begin{bmatrix}
1 & i\Omega_c \\
-i\Omega_c & 1
\end{bmatrix},
\end{align}
where $\Omega_c$ is a dimensionless quantity proportional to $B$:
\begin{align} \Omega_c=\frac{\omega\omega_c}{\omega_0^2-\omega^2}=\frac{2[1+\chi_0(\omega)]}{\chi_0(\omega)}\frac{V_dB}{\beta},
\end{align}
$\omega_0=\sqrt{k_0/m_e}$ is the angular frequency of harmonic vibration, and $\omega_c=eB/m_e$ is the cyclotron frequency.
The Verdet constant is $V_d(\omega)=\frac{e\omega}{2m_ec}\frac{d{n}}{d\omega}$.
$\chi_0(\omega)$ is the first-order susceptibility without an external magnetic field:
\begin{align}\label{chi0}
\chi_0(\omega)=\frac{{Ne^2}/{m_e\epsilon_0}}
{\omega_0^2-\omega^2},
\end{align}
where $N$ and $\epsilon_0$ are the effective density of electron number and permittivity of vacuum.
The refractive index without a magnetic field is $n(\omega)=\sqrt{1+\chi_0(\omega)}$, and furthermore the propagation constant is $\beta=n(\omega)\omega/c$, where $c$ is the speed of light in vacuum.
The tensor $\chi^{(1)}$ in Eq. (\ref{chi1}) possesses rotational invariance,
\begin{align}
  \chi^{(1)}_{i'j'}={\rm R}^{-1}_{i'i}\chi^{(1)}_{ij}{\rm R}_{jj'},\quad {\rm R}_{ij}=\begin{bmatrix}
                           \cos\theta & \sin\theta \\
                           -\sin\theta & \cos\theta
                         \end{bmatrix}_{ij},
\end{align}
which is an important property ensuring that the third-order susceptibility remains unchanged during the rotation of the optical fiber. Therefore, the effects of slight twisting during the coiling process can be ignored.
In current experiments of optical fiber, we always have $V_dB\ll\beta$, so we neglect the second-order term of $\Omega_c$ and obtain $\chi^{(1)}=\chi_0(\mathbf{I}-\Omega_c\hat{\sigma}_2)$, where $\hat{\sigma}_2$ is Pauli matrix.

Next, substituting equation (\ref{xy61}) into equation (\ref{move7}), retaining terms containing $e^{-3i\omega t}$, and ignoring higher-order terms of $k_2$ and $x^{(3)},y^{(3)}$, we obtain the amplitudes of the third-harmonic part $x^{(3)},y^{(3)}$ [see Eq. (\ref{x3y3}) in Appendix II], which are the nonlinear functions about $A_x$ and $A_y$.
We know that the nonlinear polarization vector has the following form:
\begin{align}\label{PNi62}		{P}_{i}^{(N)}&=\epsilon_0{\chi}^{(3)}_{ijkl}E_jE_kE_l\nonumber\\
		&=\frac{\epsilon_0}{8}{\chi}^{(3)}_{ijkl}[({A^*_j}A_kA_l+A_j{A^*_k}A_l+A_jA_k{A^*_l})e^{-i\omega t}\nonumber\\
  &\qquad+({A_j}A^*_kA^*_l+A^*_j{A_k}A^*_l+A^*_jA^*_k{A_l})e^{i\omega t}\nonumber\\
		&\qquad+A_jA_kA_le^{-3i\omega t}+A_j^*A_k^*A_l^*e^{3i\omega t}]\nonumber\\
		&\equiv\frac{1}{2}(N_i^{(1)}e^{-i\omega t}+N_i^{(1)*}e^{i\omega t}+N_i^{(3)}e^{-3i\omega t}+N_i^{(3)*}e^{3i\omega t}),
\end{align}
where $N_i^{(1)}$ and $N_i^{(3)}$ are the amplitudes of the fundamental and the third harmonic parts, respectively.
For the third harmonic part, there is a relationship between the nonlinear polarization vector and the electron displacement vector (their $i$-th components are $P_i^{(N)}$ and $r_i$, respectively): $P_i^{(N)}|_{3\omega}=-Ner_i|_{3\omega}$.
Here $N$ is the number of electrons per unit volume. From Eqs. (\ref{PNi62}) and (\ref{xy61}), we know $N_x^{(3)}=-Nex^{(3)}$ and $N_y^{(3)}=-Ney^{(3)}$.
By comparing them with equation (\ref{PNi62}), we can obtain the components of the third-order susceptibility tensor $\chi^{(3)}$:
\begin{subequations}
\begin{align}\label{chiijkl}
	&\chi^{(3)}_{xxxx}=\chi^{(3)}_{yyyy}=-\frac{\epsilon_0^3\chi_0^3(\omega)\chi_0(3\omega)}{N^3e^4}k_2,\\
	&\chi^{(3)}_{xxyy}=\chi^{(3)}_{xyyx}=\chi^{(3)}_{xyxy}=\chi^{(3)}_{yxxy}=\chi^{(3)}_{yyxx}=\chi^{(3)}_{yxyx}=\frac{1}{3}\chi^{(3)}_{xxxx},\\
	&\chi^{(3)}_{xyyy}=-\chi^{(3)}_{yxxx}=\frac{4i\chi_0(3\omega)}{\chi_0(\sqrt{3}\omega)}\Omega_c\chi^{(3)}_{xxxx},\\
	&\chi^{(3)}_{xxxy}=\chi^{(3)}_{xxyx}=\chi^{(3)}_{xyxx}=\frac{1}{3}\chi^{(3)}_{xyyy},\\ &\chi^{(3)}_{yxyy}=\chi^{(3)}_{yyxy}=\chi^{(3)}_{yyyx}=-\frac{1}{3}\chi^{(3)}_{xyyy}.
\end{align}
\end{subequations}
This tensor $\chi^{(3)}$ also possesses rotational invariance, $
  \chi^{(3)}_{i'j'k'l'}={\rm R}^{-1}_{i'i}{\rm R}^{-1}_{j'j}\chi^{(3)}_{ijkl}{\rm R}_{kk'}{\rm R}_{ll'}$,
to ensure that it remains unchanged during the rotation of the optical fiber in tangential direction.
In the absence of the external magnetic field ($\Omega_c=0$),  the last eight elements of this tensor will become $0$, and the derived $\chi^{(3)}$ is the same as the result in Ref. \cite{Boyd-book}.

\subsection{Nonlinear optical propagation equation}

We consider a helically wound single-mode nonlinear fiber, as shown in Fig. \ref{pic-model}.
Due to the polarization vector $\vec{P}=\vec{P}^{(L)}+\vec{P}^{(N)}$, in the Frenet coordinate system, the propagation equation of electric field of the light is
\begin{align}\label{helm44}
		\frac{\partial^2 \vec{E}(s,t)}{\partial s^2}=\frac{1}{c^2}\frac{\partial^2 \vec{E}(s,t)}{\partial t^2}+\mu_0 \frac{\partial^2 \vec{P}^{(L)}(s,t)}{\partial t^2}+\mu_0 \frac{\partial^2 \vec{P}^{(N)}(s,t)}{\partial t^2},
\end{align}
where the electric field vector $\vec{E}(s,t)=E_n(s,t)\hat{n}(s)+E_b(s,t)\hat{b}(s)$, and the propagation distance is represented by the arc length $s$.
The specific calculation process of Eq. (\ref{helm44}) can be seen in Appendix I.
When we input a continuous wave, the electric field vector can be written as
\begin{eqnarray}
	\begin{split}\label{eee}
		\vec{E}(s,t)=\frac{1}{2}[\vec{A}(s)\,e^{i({\beta} s-\omega t)}+\vec{A}^*(s)\,e^{-i({\beta} s-\omega t)}],
	\end{split}
\end{eqnarray}
where the propagation constant is $\beta=n(\omega)\omega/c$ and the complex amplitude vector of the electric field is $\vec{A}(s)=A_n(s)\hat{n}(s)+A_b(s)\hat{b}(s)$.

The linear polarization vector is $\vec{P}^{(L)}(s,t)=\epsilon_0\chi^{(1)}\vec{E}(s,t)$, where the first-order susceptibility has the expression (\ref{chi1}); the nonlinear one is $\vec{P}^{(N)}(s,t)=\epsilon_0{\chi}^{(3)}\vdots\,
\vec{E}(s,t)\vec{E}(s,t)\vec{E}(s,t)$, where the third-order susceptibility has the expression (\ref{chiijkl}).
Ignoring the third-harmonic waves in the fiber, we recall the nonlinear polarization vector (\ref{PNi62}) and write
\begin{align}\label{PNpp} \vec{P}^{(N)}(s)=\frac{1}{2}[\vec{N}^{(1)}(s)e^{i({\beta} s-\omega t)}+\vec{N}^{*(1)}(s)e^{-i({\beta} s-\omega t)}].
\end{align}
where $\vec{N}^{(1)}(s)=N_n^{(1)}(s)\hat{n}(s)+N_b^{(1)}(s)
\hat{b}(s)$
and
\begin{eqnarray}
	\begin{split}\label{Ni} {N}_i^{(1)}=\frac{\epsilon_0}{4}{\chi}^{(3)}_{ijkl}({A^*_j}
A_kA_l+A_j{A^*_k}A_l+A_jA_k{A^*_l}).
	\end{split}
\end{eqnarray}
Furthermore, substituting the expression (\ref{chiijkl}) of $\chi^{(3)}$ into it, one can obtain
\begin{subequations}\label{Nxy612}
\begin{align}
&N_n^{(1)}=\frac{3\epsilon_0}{4}{\chi}^{(3)}_{xxxx}\Big(h_1+iC_B\frac{V_dB}{\beta}h_2\Big),\\
&N_b^{(1)}=\frac{3\epsilon_0}{4}{\chi}^{(3)}_{xxxx}\Big(h_2-iC_B\frac{V_dB}{\beta}h_1\Big).
\end{align}
\end{subequations}
where
\begin{subequations}
\begin{align}
&h_1=(|{A_n}|^2+\frac{2}{3}|{A_b}|^2)A_n+\frac{1}{3}{A^*_n}A_b^2,\\
&h_2=(\frac{2}{3}|{A_n}|^2+|{A_b}|^2)A_b+\frac{1}{3}A_n^2{A^*_b},
\end{align}
\end{subequations}
and $C_B$ is a dimensionless function about light frequency:
\begin{align}\label{CB}
C_B=\frac{8\chi_0(3\omega)[1+\chi_0(\omega)]}
{\chi_0(\omega)\chi_0(\sqrt{3}\omega)}.
\end{align}

\begin{table*}[htbp]
	\caption{Rotation angle $\theta/l$ of light in a nonlinear fiber (per unit evolution distance)}
	\label{tab-1} \begin{tabular}{p{2.5cm}<{\centering}|p{3.5cm}<{\centering}
|p{4.5cm}<{\centering}|p{5.5cm}<{\centering}}
		\hline
		\hline
		 & Straight fiber with an external magnetic & Helically wound fiber without an external magnetic field & Helically wound fiber with an external magnetic field  \\
		\hline
		Linearly polarized light & $(1-C_B\frac{\sigma' A^2}{\beta})V_dB$ & $-(1+\frac{\sigma' A^2}{\beta})\tau$ & $(1-C_B\frac{\sigma' A^2}{\beta})V_dB-(1+\frac{\sigma' A^2}{\beta})\tau$ \\
		\hline
		Elliptically polarized light & $(1-C_B\frac{\sigma' A^2}{\beta})V_dB+\frac{1}{3} {\sigma'}A^2\sin 2\chi$ & $-(1+\frac{\sigma' A^2}{\beta})\tau+\frac{1}{3} {\sigma'}A^2\sin 2\chi$ & $(1-C_B\frac{\sigma' A^2}{\beta})V_dB-(1+\frac{\sigma' A^2}{\beta})\tau+\frac{1}{3} {\sigma'}A^2\sin 2\chi$ \\
		\hline
		\hline
	\end{tabular}
\end{table*}

Next, we substitute the expressions of $\vec{E}$ (\ref{eee}), $\vec{P}^{(L)}$, and $\vec{P}^{(N)}$ (\ref{PNpp}) into Eq. (\ref{helm44}) to obtain:
\begin{align}\label{all4}
2i\beta\frac{d\vec{A}}{d s}+\frac{d^2\vec{A}}{d s^2}-2\beta V_d B\hat{\sigma}_2\vec{A}+\mu_0\omega^2\vec{N}^{(1)}=0.
\end{align}
Due to $\frac{d\hat{\alpha}}{ds}=\kappa\hat{n}$, $\frac{d\hat{n}}{ds}=-\kappa\hat{\alpha}+\tau\hat{b}$,
$\frac{d\hat{b}}{ds}=-\tau\hat{n}$, considering the slowly varying envelope approximation and neglecting the terms of $\frac{d^2A_i}{ds^2}$ in the above equations, we can obtain
\begin{subequations}
\begin{align}
	&\frac{dA_n}{d s}=\frac{i\beta(\tau^2-\kappa^2-2V_dB\tau)}{2(\beta^2-\tau^2)}A_n
+\frac{2\beta^2\tau-\tau^3-2V_dB\beta^2}{2(\beta^2-\tau^2)}A_b\nonumber\\
&\qquad +\frac{i\mu_0\omega^2}{2(\beta^2-\tau^2)}(\beta N_n-i\tau N_b),\\
	&\frac{dA_b}{d s}=\frac{\tau(\kappa^2+\tau^2-2\beta^2)+2V_dB\beta^2}{2(\beta^2-\tau^2)}A_n+
\frac{i\beta\tau(\tau-2V_dB)}{2(\beta^2-\tau^2)}A_b\nonumber\\
&\qquad +\frac{i\mu_0\omega^2}{2(\beta^2-\tau^2)}(i\tau N_n+\beta N_b).
\end{align}
\end{subequations}
Then, we consider the practical condition $\tau,\kappa,V_dB\ll\beta$, neglect the quadratic terms of the small quantities $\tau/\beta$, $\kappa/\beta$, and $V_dB/\beta$, and substitute the expressions (\ref{Nxy612}) of $N_n$ and $N_b$ into the above equations.
Due to practical reasons, in Eq. (\ref{Nxy612}), $A_x$ and $A_y$ are expressed in units of electric field ($\text{V/m}$).
We introduce $\vec{A}'$ such that $|\vec{A}'|^2$ to represent optical power:
\begin{align}
	|\vec{A}'|^2=\frac{1}{2}nc\epsilon_0S|\vec{A}|^2,
\end{align}
where $S$ is the effective cross-sectional area of the fiber core.
After substituting the above relation, for simplicity, we omit the prime notation from $A'_x$ and $A'_y$.
Then we can obtain:
\begin{align}\label{anab}
	i\frac{d\vec{A}}{d s}=\Big[\hat{K}+(V_dB-\tau)\hat{\sigma}_2&+\frac{1}{\beta}
(\tau+C_BV_dB)\hat{\sigma}_2\hat{h}'\nonumber\\
&+(1+\frac{V_dB\tau}{\beta^2}C_B)\hat{h}'\Big]\vec{A},
\end{align}
where $\hat{h}'$ is the Homitonian in a straight nonlinear fiber when the magnetic field is absent:
\begin{align}
	\hat{h}'=-\sigma'\begin{bmatrix}
		|{A_n}|^2+\frac{2}{3}|{A_b}|^2 & \frac{1}{3} A_n^*A_b \\
		\frac{1}{3} A_nA_b^* & \frac{2}{3}|{A_n}|^2+|{A_b}|^2 \\
	\end{bmatrix},
\end{align}
and $\sigma'=\frac{3{\chi}^{(3)}_{xxxx}\omega}{4n^2c^2\epsilon_0S}$.
The matrix $\hat{K}$ is
\begin{align}
	\hat{K}=\begin{bmatrix}
		\frac{\kappa^2}{2\beta} & 0 \\
		0 & 0 \\
	\end{bmatrix}+\frac{2V_dB\tau-\tau^2}{2\beta}\begin{bmatrix}
                                               1 & 0 \\
                                               0 & 1
                                             \end{bmatrix}.
\end{align}
The first part arises from the linear birefringence induced by curvature $\kappa$, which can be neglected by considering $2\pi R\ll H$ such that $\kappa \ll \tau$.
The second part represents the change in propagation constants, but due to $\Delta\beta = \frac{2V_dB\tau - \tau^2}{2\beta} \ll \beta$, it can be ignored.
Therefore, we can ignore $\hat{K}$ and the quadratic small quantity $\frac{V_dB\tau}{\beta^2}$, yielding:
\begin{align}\label{anab1}
	i\frac{d\vec{A}}{d s}=\Big[(V_dB-\tau)\hat{\sigma}_2+\frac{1}{\beta}(\tau+C_BV_dB)
\hat{\sigma}_2\hat{h}'+\hat{h}'\Big]\vec{A}.
\end{align}
This is the evolution equation of the amplitude vector $\vec{A}$ with the arc length $s$.

Three interesting situations will be discussed as follows.
In the absence of the nonlinear effect ($\sigma'=0$), Eq. (\ref{anab1}) will become $i\partial_s\vec{A}=(V_dB-\tau)\hat{\sigma}_2\vec{A}$, where the equivalence between $V_dB$ and $\tau$ indicates that the torsion $\tau$ can be viewed as an effective magnetic field \cite{Berry-1987,Ross-1984,Tabor-1969,Smith-1978,Rashleigh-1983}, i.e., the so-called geometry-induced gauge fields \cite{Guinea-2010,Zhang-2014,Wang-2014,Tan-2021,Wang-2022}.
In a straight fiber without the external magnetic field ($B=0$ and $\tau=0$), Eq. (\ref{anab1}) will become $i\partial_s\vec{A}=\hat{h}'\vec{A}$, which describes the evolution of a continuous wave in an isotropic nonlinear fiber \cite{Crosignani-1985,Trillo-1986,Akhmediev-1994,Barad-1997}.
In a helically wound fiber with an external magnetic field, Eq. (\ref{anab1}) describes the evolution of a continuous wave under the combined impact of the magnetic field, helical geometry, and the nonlinear effect.
In this equation, the term $\frac{1}{\beta}(\tau+C_BV_dB)
\hat{\sigma}_2\hat{h}'$ manifests an interesting interplay between the three effects, providing a comparable correction of rotation angle in the cases with intense lights.

\section{Nonlinear Faraday effect in a helically wound fiber}

\subsection{Analytic expressions of the nonlinear Faraday angle}

We analyze the evolution of polarization states in this system from another perspective  \cite{Barad-1997}.
Introducing the Stokes vector $\vec{S}=S_x\hat{e}_x+S_y\hat{e}_y+S_z\hat{e}_z$, where the components are defined as:
\begin{align}\label{s1s2s3}
	&S_x=|A_x|^2-|A_y|^2,\; S_y= A_x^* A_y+ A_x A_y^*,\; S_z=i( A_x A_y^*- A_x^* A_y).
\end{align}
It's noted that the amplitude of this vector is conserved: $|\vec{S}|=|A_x|^2+|A_y|^2=A^2$, i.e., it equals to the light power.
Substituting Eq. (\ref{s1s2s3}) into Eq. (\ref{anab1}), we obtain:
\begin{align}\label{sds}
	\frac{d\vec{S}}{dz}=-2\Big[V_dB-\tau-\frac{\sigma' A^2}{\beta}(\tau+C_BV_dB)+\frac{1}{3} {\sigma'}S_z\Big]\vec{S}\times\hat{e}_z.
\end{align}
It describes the rotation of vector $\vec{S}$ around $\hat{e}_z$ axis, which indicates that $S_z$ is conserved. The rotation solution to this motion equation is:
\begin{align}\label{solveS}
\vec{S}&=(S_x,S_y,S_z)=A^2\cos 2\chi[\cos (2\Omega z),\, \sin (2\Omega z),\, \tan 2\chi],
\end{align}
where
\begin{align}
	&\Omega=V_dB-\tau-\frac{\sigma' A^2}{\beta}(\tau+C_BV_dB)+\frac{1}{3} {\sigma'}A^2\sin 2\chi.
\end{align}
$\chi$ represents the elliptic angle, whose zero or non-zero values correspond to linearly or elliptically polarized states, respectively.

The solution (\ref{solveS}) describes a rotation process of different polarized states in the Poincar$\rm \acute{e}$ sphere.
Its rotation frequency of $\vec{S}$ in the sphere is twice the rotation frequency of the light's polarization in the real space of $n$-$b$.
Therefore, we can derive the rotation angle per unit length for linearly polarized light ($\chi=0$) as:
\begin{align}\label{thetaLP}
	{\theta}/{l}=\Omega=\Big(1-C_B\frac{\sigma' A^2}{\beta}\Big)V_dB-\Big(1+\frac{\sigma' A^2}{\beta}\Big)\tau,
\end{align}
and for elliptically polarized light ($\chi\neq 0$) as:
\begin{align}\label{thetaEP}
	{\theta}/{l}=\Omega=\Big(1-&C_B\frac{\sigma' A^2}{\beta}\Big)V_dB-\Big(1+\frac{\sigma' A^2}{\beta}\Big)\tau+\frac{1}{3} {\sigma'}A^2\sin 2\chi,
\end{align}
where $\theta$ is total rotation angle of light and $l$ is total propagation distance,.
The dimensionless quantity $C_B$ has the expression (\ref{CB}), which measures the impact of nonlinear effect on the Faraday rotation angle.
For a nonlinear fiber, the rotation angle $\theta/l$ in different situations is illustrated in Tab. \ref{tab-1}.

\subsection{Some discussions}

We firstly focus on the rotation of linearly polarized light, as shown in the second row of Tab. \ref{tab-1}.
In a straight fiber with an external magnetic field, nonlinear effects will exert a correction on the Verdet constant, resulting in a modified Verdet constant of $V_d^{(N)}=(1-C_B\frac{\sigma' A^2}{\beta})V_d$.
Similarly, in a helically wound fiber without an external magnetic field, nonlinear effects will also influence the rotation angle in the Faraday-like effect, equivalent to exerting a correction to the torsion of the fiber, resulting in a modified torsion of $\tau^{(N)}=(1+\frac{\sigma' A^2}{\beta})\tau$. When considering both the winding of the fiber and the external magnetic field simultaneously, the obtained rotation angle is a linear combination of the first two results.

Then, we shift our focus to elliptically polarized light, as shown in the third row of Tab. \ref{tab-1}.
In the three cases mentioned above, the only difference between the rotation angles of linearly and elliptically polarized light is the emergence of an additional term $\frac{1}{3} {\sigma'}A^2\sin 2\chi$.
It indicates that the rotation angle of an elliptically polarized light will also depend on its own ellipticity angle $\chi$.
In a straight fiber without an external magnetic field, due to the presence of this term, an elliptically polarized light also exhibits rotation.
This kind of intensity-induced rotation is fundamentally different from the Faraday and Faraday-like rotations, and has been earlier studied in various systems including nonlinear fibers \cite{Crosignani-1985,Maker-1964,Agrawal-book}.

Among these situations in Tab. \ref{tab-1}, an interesting one is a helically wound optical fiber without an external magnetic field, where the rotation angle is $\theta/l=-(1+\frac{\sigma' A^2}{\beta})\tau$.
In the absence of the nonlinearity, the torsion geometry serves as the magnetic field, and is closely related to the celebrated Berry phase, as discussed by Ref. \cite{Berry-1987}.
The geometry induced gauge field is rather interesting and constantly attract much attention \cite{Wang-2014,Wang-2022}.
In the presence of the nonlinearity, it is seen that the rotation angle is due to the interplay between the geometry and nonlinearity, which indicates the emergence of a nonlinear Berry phase \cite{Liu-2010,Liu-book}.

\subsection{The observations of nonlinear Faraday rotation}

\textbf{Parameter Settings}\quad
We use the optical fibers made of $\rm As_2S_3$ as the medium for light propagation, which is a type of chalcogenide fiber. Compared to traditional $\rm SiO_2$ fibers, this type of fiber exhibits larger nonlinear refractive index and Verdet constant, making it suitable for observing nonlinear and magnetic field-induced effects. The nonlinear refractive index \cite{Sanghera-2008} and the Verdet constant \cite{Ruan-2005} of $\rm As_2S_3$ fibers have been measured in previous experiments, and here we will refer to the parameters provided in the two experimental works.

Firstly, for the nonlinearity coefficient $k_2$, we can estimate it using the following formula \cite{Boyd-book}:
\begin{align}\label{eq-k2}
  k_2=-\frac{k_0}{d^2}=-\frac{m_e\omega_0^2}{d^2},
\end{align}
where $\omega_0$ is the lowest resonant frequency of the atom. $d$ is a typical value for the atomic scale, which can be taken as the Bohr radius: $d\approx a_0=\frac{4\pi\epsilon_0\hbar^2}{m_ee^2}$.

In the non-resonant limit ($\omega\ll\omega_0$), from Eq. (\ref{chi0}), the first-order polarization without an external magnetic field can be expressed as:
\begin{align}\label{eq-chi0lim}
  \chi_0\approx\frac{Ne^2}{m_e\epsilon_0\omega_0^2}.
\end{align}
Meanwhile, from Eq. (\ref{chiijkl}), the typical component of the third-order polarization can be written as:
\begin{align}\label{eq-chixlim}
  \chi_{xxxx}&=-\frac{\epsilon_0^3\chi_0^3(\omega)\chi_0(3\omega)}
  {N^3e^4}k_2\approx \frac{Ne^4}{m_e^3\epsilon_0d^2\omega_0^6}.
\end{align}
The Verdet constant can be written as:
\begin{align}\label{eq-vdlim}
   V_d=\frac{e\omega}{2m_ec}\frac{d{n}}{d\omega}&=\frac{N\epsilon_0e^3\omega^2}{2c\sqrt{\epsilon_0
      m_e(\omega_0^2-\omega^2)+Ne^2}}\nonumber\\
      &\approx \frac{N\epsilon_0e^3\omega^2}{2c\sqrt{\epsilon_0
      m_e\omega_0^2+Ne^2}}
\end{align}
It has been experimentally measured that the lowest bandgap of $\rm As_2S_3$ is $E_g=2.38\, {\rm eV}$, thus the lowest resonant frequency is $\omega_0=E_g/\hbar=3.6\times 10^{15}\,{\rm Hz}$. Based on it, the following are the calculations of some parameters:
\begin{itemize}
  \item Density of $\rm As_2S_3$ fiber: $\rho=3.46\,{\rm g/cm^3}$, molecular mass: $M=246.04\times 1.661\times 10^{-27}\,{\rm kg}$, thus the number density of molecules: $N=\rho/M=8.47\times 10^{27}/{\rm m^3}$.
  \item Refractive index: $n=\sqrt{1+\chi_0}= 1.75$, experimental result: $\bar{n}=2.4$.
  \item Typical component of the third-order polarization: $\chi_{xxxx}= 1.33\times 10^{-19}\,{\rm m^2/V^2}$, experimental result: $\bar{\chi}_{xxxx}=4.1\times 10^{-19}\,{\rm m^2/V^2}$.
  \item Nonlinear refractive index: $n_2=\frac{3}{4n^2\epsilon_0 c}\chi^{(3)}_{xxxx}=1.23\times 10^{-17}\,{\rm m^2/W}$, experimental result: $\bar{n}_2=2\times 10^{-17}\,{\rm m^2/W}$.
  \item Verdet constant: $V_d=39.2\,{\rm /T/m}$, experimental result: $\bar{V}_d=14.5{\rm /T/m}$.
\end{itemize}

\textbf{Nonlinear Correction to Verdet Constant}\quad
Here, we focus on the situation of a straight fiber to discuss the correction induced by the third-order nonlinear effect for Verdet constant.
From Eqs. (\ref{thetaLP}) and (\ref{thetaEP}), in the presence of an external magnetic field and considering third-order nonlinear effects, the rotation angle per unit distance of linearly polarized light ($\chi=0$) in a straight fiber is given by
\begin{align}
	\theta/l=\Omega=\Big(1-\frac{\sigma' A^2}{\beta}C_B\Big)V_dB,
\end{align}
and the one of elliptically polarized light ($\chi\neq 0$) is given by
\begin{align}
	\theta/l=\Omega=\Big(1-\frac{\sigma' A^2}{\beta}C_B\Big)V_dB+\frac{1}{3} {\sigma'}A^2\sin 2\chi.
\end{align}
$C_B$ is an important parameter, measuring the influence of third-order nonlinear effects on the Faraday rotation angle.
Interestingly, it can considered as a nonlinear correction to the Verdet constant, i.e., in the nonlinear case, the Verdet constant becomes
\begin{align}\label{eq-vdnl}
  V_d^{(N)}=\Big(1-\frac{\sigma' A^2}{\beta}C_B\Big)V_d.
\end{align}

In the non-resonant limit, from Eq. (\ref{CB}), the parameter $C_B$ can be approximated as:
\begin{align}\label{CB2}
C_B\approx \frac{8(1+\chi_0)}{\chi_0}=\frac{8n^2}{n^2-1}=8\Big(1+\frac{m_e\epsilon_0\omega_0^2}
{Ne^2}\Big).
\end{align}
Using the practical refractive index $\bar{n}$, the value of $C_B$  in $\rm As_2S_3$ fibers is $C_B=8\bar{n}^2/(\bar{n}^2-1)=11.33$.
In materials with refractive index greater than 1, the value of $C_B$ decreases with increasing refractive index.

According to Ref. \cite{Sanghera-2008}, we take the diameter of the fiber core as $d_c=4.2\,{\rm \mu m}$, core cross-sectional area as $S=13.85\,{\rm \mu m^2}$, incident wavelength as $\lambda=1550\,{\rm nm}$, and propagation constant as $\beta=2\pi/\lambda=5.87\times 10^6\,{\rm m^{-1}}$.
Assuming the light has approximately uniform distribution in the radial direction, we can obtain the nonlinear coefficient $\sigma'$ as
\begin{align}\label{eq-sigma}
  \sigma' = \frac{2\pi \bar{n}_2}{\lambda S}=5.85\,{\rm /W/m}=3.6\,{\rm /W/m}.
\end{align}
For a linearly polarized light, according to Eq. (\ref{thetaLP}), we need $C_B\sigma' A^2/\beta$ to be comparable to 1 to ensure the correction part is significant.
Here, we set the condition as $C_B\sigma' A^2/\beta>0.01$, which results in a required optical power: $A^2>0.01\beta/\sigma'C_B=611.6\,{\rm W}$, thus the required incident intensity is $I=A^2/S>4.42\times 10^{9}\,{\rm W/cm^2}$.

It is worthy noting that the nonlinear coefficient $\sigma'$ is also dependent on the refractive index $n$, thus the realistic correction of nonlinear effects for Verdet constant is measured by $\sigma' C_B$.
Its expression is
\begin{align}\label{sCB}
  \sigma'C_B=\frac{3\epsilon_0\omega}{4c^2N^2e^2d^2S}
  \,(n^2-1)^2.
\end{align}
In materials with refractive index greater than 1, the value of $\sigma' C_B$ increases with increasing $n$.
Therefore, to exhibit a greater correction part of the rotation angle, it is preferable to choose a material with a higher refractive index.
\\

\textbf{Observation of Rotation Angle in a Hellically Wound Fiber}\quad
To better observe the influence of nonlinear effects on the rotation angle, we choose to observe in a wound fiber system
In this system, by selecting an appropriate geometric configuration, the effects of external magnetic field and curvature cancel each other out, i.e., $\tau=V_d B$.
The deflection angle per unit distance of linearly polarized light becomes
\begin{align}\label{thetawound}
	\theta/l=\Omega=-(1+C_B)\sigma' A^2\frac{\tau}{\beta}.
\end{align}
It means that the rotation angle is proportional to the light power, thus provides the possibility of measuring the power of a strong light and the nonlinear coefficient.
To satisfy $\kappa\ll \tau$, taking the helix pitch of the fiber winding as $H=19\,{\rm cm}$, the length of one turn of the fiber winding as $L=20\,{\rm cm}$, the winding radius is $R=\sqrt{L^2-H^2}/2\pi=1\,{\rm cm}$, thus the curvature is $\kappa=\frac{(2\pi)^2R}{L^2}=9.8\,{\rm /m}$, and the torsion rate is $\tau=\frac{2\pi H}{L^2}= 29.8\,{\rm /m}$.
Considering that the required tangential field intensity is $B={\tau}/{V_d}$, the magnetic field intensity in the $z$ direction is
\begin{align}\label{eq-B}
 B_z=\frac{L}{H}B=\frac{2\pi}{LV_d}=2.17\,{\rm T}.
\end{align}
According to Eq. (\ref{thetawound}), to observe a rotation angle whose absolute value is greater than $\theta_m=0.01\times 2\pi\,{\rm rad}$, the optical power required is
\begin{align}
	A^2>\frac{\beta\theta_m}{(1+C_B)\sigma' \tau L}\approx 591.6\,{\rm W}.
\end{align}
Thus, the required incident intensity is $I=A^2/S>4.27\times 10^{9}\,{\rm W/cm^2}$.

\section{Conclusion}

In summary, we analyze the influence of third-order nonlinear optical effects on the Faraday rotation and find that the nonlinearity might modify the Verdet constant as well as the geometric torsion, leading to additional terms in the expressions of the rotation angle for both linearly and elliptically polarized lights.
Possible observations of our findings in $\rm As_2S_3$ fibers are discussed.
Our theoretical results bring a prospect for the observations of the light-intensity-dependent Faraday rotation and the nonlinear Berry phase, and might have potential applications in various areas such as the measurements of strong electromagnetic fields.

\section*{Acknowledgement}
This work was supported by NSAF (No.U2330401) and National Natural Science Foundation of China (No. 12247110).

\appendix
\setcounter{equation}{0}
\renewcommand\theequation{A\arabic{equation}}

\section*{Appendix I: Propagation equation of light in the Frenet coordinate system}
\label{app2}

For the self-consistence of the paper, here, we present the detailed deductions of propagation equation of light in the Frenet coordinate system.
First, we investigate the propagation equation of light waves in the laboratory coordinate system ($\hat{x},\hat{y},\hat{z}$).
For light waves propagating in optical fibers, the evolution of their electric and magnetic field components satisfies the Maxwell's equations:
\begin{align}\label{max}
		&\nabla\times\vec{E}(\vec{r},t)=-\frac{\partial \vec{B}(\vec{r},t)}{\partial t}, \quad \nabla\times\vec{H}(\vec{r},t)=\frac{\partial \vec{D}(\vec{r},t)}{\partial t}, \nonumber\\
		&\qquad \qquad \nabla\cdot\vec{D}(\vec{r},t)=0,\quad \nabla\cdot\vec{B}(\vec{r},t)=0.
\end{align}
Substituting $\vec{D}(\vec{r},t)=\epsilon_0\vec{E}(\vec{r},t)+\vec{P}(\vec{r},t)$ and $\vec{B}(\vec{r},t)=\mu_0\vec{H}(\vec{r},t)$, we obtain
\begin{eqnarray}
	\begin{split}\label{helm1}
		\nabla[\nabla\cdot\vec{E}(\vec{r},t)]-\nabla^2\vec{E}(\vec{r},t)=-\frac{1}{c^2}\,\frac{\partial^2 \vec{E}(\vec{r},t)}{\partial t^2}-\mu_0 \frac{\partial^2 \vec{P}(\vec{r},t)}{\partial t^2},
	\end{split}
\end{eqnarray}
where $\vec{P}$ is the polarization vector, and $c$ is the speed of light in vacuum. Here we consider weakly guiding optical fibers commonly used in communication, with a small difference in refractive index between the core and cladding, so the rate of change of the relative dielectric tensor in the transverse direction is small. Thus, we have $\nabla\cdot\vec{E}(\vec{r},t)\approx\frac{\epsilon^{-1}}{\epsilon_0}\nabla\cdot\vec{D}(\vec{r},t)=0$. Substituting this into Equation (\ref{helm1}) eliminates the first term, resulting in the transmission equation for the electric field in the laboratory coordinate system (subscript $L$):
\begin{eqnarray}
	\begin{split}\label{helm2}
		\nabla_L^2\vec{E}(\vec{r}_L,t)=\frac{1}{c^2}\,\frac{\partial^2 \vec{E}(\vec{r}_L,t)}{\partial t^2}+\mu_0 \frac{\partial^2 \vec{P}(\vec{r}_L,t)}{\partial t^2}.
	\end{split}
\end{eqnarray}
Here, $\vec{r}_L=[x,y,z]^{\rm T}$ is the position vector in the laboratory coordinate system, and $\nabla^2_L=\frac{\partial^2}{\partial x^2}+\frac{\partial^2}{\partial y^2}+\frac{\partial^2}{\partial z^2}$ is the three-dimensional Laplacian operator in the laboratory coordinate system.

If the coordinate origin is set at the center of the left end of the cylinder, in the laboratory coordinate system, the position curve of the fiber can be expressed as
\begin{eqnarray}
	\begin{split}\label{r}
		\vec{f}_L(s)=\begin{bmatrix}
			f_X(s) \\
			f_Y(s) \\
			f_Z(s)
		\end{bmatrix}=\begin{bmatrix}
			R\cos(\frac{2\pi}{L}s) \\
			R\sin(\frac{2\pi}{L}s) \\
			\frac{H}{L}s
		\end{bmatrix},
	\end{split}
\end{eqnarray}
where \(s\) is the arc length of the fiber curve, \(R\) is the radius of the cylinder, \(H\) is the pitch of the fiber (i.e., the distance the light moves in the \(z\) direction after one winding cycle), and \(L=\sqrt{(2\pi R)^2+H^2}\) is the length of the fiber within one winding cycle.
Therefore, the expressions for the Frenet coordinate base vectors in the laboratory coordinate system are obtained as follows:
\begin{subequations}
\begin{align} &\hat{\alpha}_L=\frac{\partial_s\vec{f}_L}{|\partial_s\vec{f}_L|}=\begin{bmatrix}
			-\frac{2\pi R}{L}\sin(\frac{2\pi}{L}s) \\
			\frac{2\pi R}{L}\cos(\frac{2\pi}{L}s) \\
			\frac{H}{L}
		\end{bmatrix},\label{eq-alpha}\\ &\hat{n}_L=\frac{\partial_s^2\vec{f}_L}{|\partial_s^2\vec{f}_L|}=\begin{bmatrix}
			-\cos(\frac{2\pi}{L}s) \\
			-\sin(\frac{2\pi}{L}s) \\
			0
		\end{bmatrix}, \\
&\hat{b}_L=\hat{\alpha}_L\times\hat{n}_L=\begin{bmatrix}
			\frac{H}{L}\sin(\frac{2\pi}{L}s) \\
			-\frac{H}{L}\cos(\frac{2\pi}{L}s) \\
			\frac{2\pi R}{L}
		\end{bmatrix}.
\end{align}
\end{subequations}
as well as the expressions for the curvature and torsion of the fiber:
\begin{subequations}
\begin{align}\label{kt} &\kappa=\Big|\frac{d^2\vec{f}_L}{ds^2}\Big|=\frac{(2\pi)^2R}{L^2},\\ &\tau=\frac{d\vec{f}_L}{ds}\cdot\Big[\frac{d^2\vec{f}_L}{ds^2}\times\frac{d^3\vec{f}_L}{ds^3}\Big]/\Big|\frac{d^2\vec{f}_L}{ds^2}\Big|^2=\frac{2\pi H}{L^2}.
\end{align}
\end{subequations}
The transformation of the position vector in the laboratory coordinate system and the Frenet coordinate system is given by
\begin{eqnarray}
	\begin{split}
		&\vec{r}_L=\begin{bmatrix}
			\hat{n}_L & \hat{b}_L & \hat{\alpha}_L
		\end{bmatrix}\vec{r}_F\equiv{\mathbf M}(s)\,\vec{r}_F,
	\end{split}
\end{eqnarray}
where
\begin{eqnarray}\label{eq-Mmatrix}
	\begin{split}
		\mathbf{M}(s)=\begin{bmatrix}
			-\cos(\frac{2\pi}{L}s) & \frac{H}{L}\sin(\frac{2\pi}{L}s) & -\frac{2\pi R}{L}\sin(\frac{2\pi}{L}s) \\
			-\sin(\frac{2\pi}{L}s) & -\frac{H}{L}\cos(\frac{2\pi}{L}s) & \frac{2\pi R}{L}\cos(\frac{2\pi}{L}s) \\
			0 & \frac{2\pi R}{L} & \frac{H}{L}
		\end{bmatrix}.
	\end{split}
\end{eqnarray}



For convenience in calculation, let us rewrite the above equation in tensor form:
\begin{eqnarray}
	\begin{split}
		x_i^{(L)}={\rm M}'_{ij}\,x_j^{(F)},
	\end{split}
\end{eqnarray}
where the superscripts $(L)$ and $(F)$ correspond to the laboratory and Frenet coordinate systems, respectively.
Taking the differential of both sides, we obtain
\begin{eqnarray}
	\begin{split}\label{xir}
		dx_i^{(L)}={\rm M}'_{ij}\,dx_j^{(F)}+x_j^{(F)}\frac{d{\rm M}'_{ij}}{ds}\,ds.
	\end{split}
\end{eqnarray}
We know
\begin{align}
	\frac{d\alpha}{d s}=\frac{d\vec{f}_L(s)}{d s}\cdot\hat{\alpha}_L(s)=1.
\end{align}
Thus, we can replace $ds$ in Eq. (\ref{xir}) with $d\alpha$:
\begin{align}
	dx_i^{(L)}={\rm M}'_{ij}\,dx_j^{(F)}+x_j^{(F)}\frac{d{\rm M}'_{ij}}{ds}\,d\alpha.
\end{align}
This yields
\begin{eqnarray}
	\begin{split}
		&\begin{bmatrix}
			{dx} \\
			{dy} \\
			{dz}
		\end{bmatrix}=\mathbf{M}_d(s)\begin{bmatrix}
			{dn} \\
			{db} \\
			{d\alpha}
		\end{bmatrix},
	\end{split}
\end{eqnarray}
where
\begin{align}
&\mathbf{M}_d(s)=\begin{bmatrix}
			-\cos(\frac{2\pi}{L}s) & \frac{H}{L}\sin(\frac{2\pi}{L}s) & -\frac{2\pi R}{L}\sin(\frac{2\pi}{L}s)+m_1 \\
			-\sin(\frac{2\pi}{L}s) & -\frac{H}{L}\cos(\frac{2\pi}{L}s) & \frac{2\pi R}{L}\cos(\frac{2\pi}{L}s)+m_2 \\
			0 & \frac{2\pi R}{L} & \frac{H}{L}\
		\end{bmatrix},\nonumber\\
	&m_1=\frac{2\pi}{L}n\sin(\frac{2\pi}{L}s)+\tau b\cos(\frac{2\pi}{L}s)-\kappa\alpha\cos(\frac{2\pi}{L}s),\nonumber\\ &m_2=-\frac{2\pi}{L}n\cos(\frac{2\pi}{L}s)+\tau b\sin(\frac{2\pi}{L}s)-\kappa\alpha\sin(\frac{2\pi}{L}s).
\end{align}
Now, the three-dimensional space interval can be expressed as
\begin{align}
	ds_{3}^2&=dx^2+dy^2+dz^2=\begin{bmatrix}
		dx & dy & dz
	\end{bmatrix}\begin{bmatrix}
	{dx} \\
	{dy} \\
	{dz}
	\end{bmatrix}\nonumber\\
&=\begin{bmatrix}
		dn & db & d\alpha
	\end{bmatrix}\mathbf{M}_d^{\rm T}\mathbf{M}_d\begin{bmatrix}
	{dn} \\
	{db} \\
	{d\alpha}
\end{bmatrix}.
\end{align}
Thus, we can obtain the covariant metric tensor:
\begin{eqnarray}
	\begin{split}
		g_{ij}=(\mathbf{M}_d^{\rm T}\mathbf{M}_d)_{ij}=\begin{bmatrix}
			1 & 0 & G_1 \\
			0 & 1 & G_2 \\
			G_1 & G_2 & G_1^2+G_2^2+G_3^2\\
		\end{bmatrix}_{ij},
	\end{split}
\end{eqnarray}
where $G_1=\kappa\alpha-\tau b, G_2=\tau n, G_3=1-\kappa n$.
Subsequently, we can obtain the contravariant metric tensor:
\begin{eqnarray}
	\begin{split}
		g^{ij}=[(\mathbf{M}_d^{\rm T}\mathbf{M}_d)^{-1}]_{ij}=\frac{1}{G_3^2}\begin{bmatrix}
			G_1^2+G_3^2 & G_1G_2 & -G_1 \\
			G_1G_2 & G_2^2+G_3^2 & -G_2 \\
			-G_1 & -G_2 & 1\\
		\end{bmatrix}_{ij},
	\end{split}
\end{eqnarray}
Therefore, in the Frenet coordinate system, the Laplacian operator is
\begin{align}
		\nabla_F^2&=\frac{1}{\sqrt{\det g}}\frac{\partial}{\partial x_F^i}(\sqrt{\det g}g^{ij}\frac{\partial}{\partial x_F^i})\nonumber\\ &=\Big(1+\frac{G_1^2}{G_3^2}\Big)\frac{\partial^2}{\partial n^2}+\Big(1+\frac{G_2^2}{G_3^2}\Big)\frac{\partial^2}{\partial b^2}+\frac{1}{G_3^2}\frac{\partial^2}{\partial \alpha^2}\\
&\quad +\frac{2G_1G_2}{G_3^2}\frac{\partial^2}{\partial n\partial b}-\frac{2G_1}{G_3^2}\frac{\partial^2}{\partial n\partial \alpha}-\frac{2G_1G_2}{G_3^2}\frac{\partial^2}{\partial b\partial \alpha}\nonumber\\
		&\quad +\Big[\kappa\Big(\frac{G_1^2}{G_3^3}-\frac{1}{G_3^2}-\frac{1}{G_3}\Big)-\tau\frac{G_2}{G_3^2}\Big]\frac{\partial}{\partial n}-\tau\frac{G_1}{G_3^3}\frac{\partial}{\partial b}
		-\kappa\frac{G_1^2}{G_3^3}\frac{\partial}{\partial \alpha}.\nonumber
\end{align}
Considering that the coordinate scale inside the fiber is much smaller than the fiber length, i.e., $n,b,\alpha\ll L$, and the curvature $\kappa$ and torsion $\tau$ are of the same order as $1/L$, we have $G_1,G_2\ll G_3\approx 1$. Therefore, we can neglect terms containing $G_1/G_3$ and $G_2/G_3$ in the above equation, yielding
\begin{eqnarray}
	\begin{split}
		\nabla_F^2&=\frac{\partial^2}{\partial n^2}+\frac{\partial^2}{\partial b^2}
		+\frac{\partial^2}{\partial \alpha^2}-2\kappa\frac{\partial}{\partial n}.
	\end{split}
\end{eqnarray}
Substituting it into Eq. (\ref{helm2}), we obtain
\begin{eqnarray}
\begin{split}\label{helm22} \nabla_F^2\vec{E}(\vec{r}_F,t)=\frac{1}{c^2}\,\frac{\partial^2 \vec{E}(\vec{r}_F,t)}{\partial t^2}+\mu_0 \frac{\partial^2 \vec{P}(\vec{r}_F,t)}{\partial t^2}.
\end{split}
\end{eqnarray}
\\
Thus, the propagation equation for light waves in the Frenet coordinate system is
\begin{align}\label{helm24}
		\Big(\frac{\partial^2}{\partial n^2}+\frac{\partial^2}{\partial b^2}
		&+\frac{\partial^2}{\partial \alpha^2}-2\kappa\frac{\partial}{\partial n}\Big)\vec{E}(\vec{r}_F,t)\nonumber\\
&=\frac{1}{c^2}\,\frac{\partial^2 \vec{E}(\vec{r}_F,t)}{\partial t^2}+\mu_0 \frac{\partial^2 \vec{P}(\vec{r}_F,t)}{\partial t^2}.
\end{align}
For a single-mode optical fiber, only the lowest mode (i.e., TEM mode) exists, where its electric field and magnetic field are perpendicular to each other and both perpendicular to the propagation direction $\hat{\alpha}$, thus it can be assumed that $E_\alpha=0$. Additionally, since the optical field is distributed near the center of the core and there is no interference from other modes, we approximately consider that the variation of the optical field in the transverse direction is very small, neglecting terms containing $\partial_n\vec{E},\partial_n^2\vec{E},\partial_b^2\vec{E}$, yielding
\begin{eqnarray}
	\begin{split}
		\frac{\partial^2\vec{E}(\alpha,t)}{\partial \alpha^2}=\frac{1}{c^2}\,\frac{\partial^2 \vec{E}(\alpha,t)}{\partial t^2}+\mu_0 \frac{\partial^2 \vec{P}(\alpha,t)}{\partial t^2}.
	\end{split}
\end{eqnarray}
Considering that $d\alpha/ds=1$ and it is not convenient to deal with in the moving local coordinate $\alpha$, we can replace the moving coordinate $\alpha$ in the above equation with the absolute coordinate $s$, obtaining
\begin{eqnarray}
	\begin{split}\label{final}
		\frac{\partial^2\vec{E}(s,t)}{\partial s^2}=\frac{1}{c^2}\,\frac{\partial^2 \vec{E}(s,t)}{\partial t^2}+\mu_0 \frac{\partial^2 \vec{P}(s,t)}{\partial t^2}.
	\end{split}
\end{eqnarray}

\section*{Appendix II: Formulas about electron oscillation}
\label{app1}

The amplitude $x^{(1)},y^{(1)}$ of fundamental part of electron's oscillation satisfies the following equations:
\begin{subequations}
\begin{align}\label{move6x1} &m_e\omega_0^2x^{(1)}-m_e\omega^2x^{(1)}+eA_x
-im_e\omega\omega_cy^{(1)}=0,\\ &m_e\omega_0^2y^{(1)}-m_e\omega^2y^{(1)}+eA_y
+im_e\omega\omega_cx^{(1)}=0.
\end{align}
\end{subequations}
The solution of $x^{(1)},y^{(1)}$ in them is
\begin{subequations}\label{x1y1}
\begin{align} &x^{(1)}=-\frac{e/m_e}{(\omega_0^2-\omega^2)^2-\omega^2\omega_c^2}[(\omega_0^2-\omega^2)A_x+i\omega\omega_cA_y],\\
&y^{(1)}=-\frac{e/m_e}{(\omega_0^2-\omega^2)^2-\omega^2\omega_c^2}[-i\omega\omega_cA_x+(\omega_0^2-\omega^2)A_y].
\end{align}
\end{subequations}
The amplitude $x^{(3)},y^{(3)}$ of third-harmonic part of electron's oscillation satisfies the following equations:
\begin{subequations}
\begin{align}\label{move6x3} &m_e(\omega_0^2-9\omega^2)x^{(1)}+\frac{k_2}{4}[(x^{(1)2}+y^{(1)2})
x^{(1)}+2(3|x^{(1)}|^2+|y^{(1)}|^2)x^{(3)}\nonumber\\
&\quad +2(x^{(1)*}y^{(1)}+x^{(1)}y^{(1)*})y^{(3)}]
-3im_e\omega\omega_cy^{(3)}=0,\\ &m_e(\omega_0^2-9\omega^2)y^{(1)}+\frac{k_2}{4}[(x^{(1)2}+y^{(1)2})
y^{(1)}+2(|x^{(1)}|^2+3|y^{(1)}|^2)y^{(3)}\nonumber\\
&\quad +2(x^{(1)*}y^{(1)}+x^{(1)}y^{(1)*})x^{(3)}]
+3im_e\omega\omega_cx^{(3)}=0.
\end{align}
\end{subequations}
Considering the approximation,
\begin{align}
  k_2|\vec{E}|^2\ll\Big(\frac{Ne}{\epsilon_0\chi_0}\Big)^2k_0= \frac{k_0^3}{e^2}.
\end{align}
the solution of $x^{(3)},y^{(3)}$ is
\begin{subequations}\label{x3y3}
\begin{align} &x^{(3)}=C_{1}A_x^3+C_{2}A_x^2A_y+C_{3}A_xA_y^2+C_{4}A_y^3,\\
&y^{(3)}=C_{5}A_x^3+C_{6}A_x^2A_y+C_{7}A_xA_y^2+C_{8}A_y^3
\end{align}
\end{subequations}
where
\begin{align}
  &C_1=\frac{e^3k_2}{4m_e^4(\omega_0^2-\omega^2)^3
  (\omega_0^2-9\omega^2)}=\frac{1}{4e^5}
  \Big(\frac{\epsilon}{N}\Big)^4\chi_0^3(\omega)\chi_0(3\omega),\nonumber\\
  &C_2=\frac{ie^3k_2(\omega_0^2-3\omega^2)\Omega_c}
  {m_e^4(\omega_0^2-\omega^2)^3
  (\omega_0^2-9\omega^2)^2}=\frac{i}{e^5}
  \Big(\frac{\epsilon}{N}\Big)^4\frac{\chi_0^3(\omega)\chi_0^2(3\omega)}
  {\chi_0(\sqrt{3}\omega)},\nonumber\\
  &C_3=C_1,\quad C_4=C_2,\quad C_5=-C_2,\nonumber\\
  &C_6=C_1,
  \quad C_7=-C_2,\quad C_8=C_1.
\end{align}

\end{document}